\documentclass[10pt]{iopart}

\usepackage{iopams}
\usepackage{graphicx}% Include figure files
\usepackage{dcolumn}% Align table columns on decimal point
\usepackage{bm}% bold math
\usepackage{wasysym}
\newcommand{\md}{\mathrm{d}}

\begin{document}

%\preprint{AIP/123-QED}

\title{Characterising superconducting filters using residual microwave background}% Force line breaks with \\
%\thanks{Footnote to title of article.}
\author{J.~S.~Lehtinen$^1$, E.~Mykk\"{a}nen$^1$, A.~Kemppinen$^1$, S.~V.~Lotkhov$^2$, D.~Golubev$^3$, and A.~J.~Manninen$^1$}

\address{$^1$VTT Technical Research Centre of Finland, Centre for Metrology MIKES, P.O. Box 1000, FI-02044
 VTT, Finland}
 \address{$^2$Physikalisch-Technische Bundesanstalt, 38116 Braunschweig, Germany}  
\address{$^3$Department of Applied Physics, Aalto University, P.O. Box 13500, FI-00076 Aalto, Finland} 
\ead{janne.lehtinen@vtt.fi}

\begin{abstract}
A normal metal -- superconductor hybrid single-electron trap with tunable barrier is utilized as a tool for spectrum analysis at the extremely low signal levels, using only well filtered cryogenic microwave background as a photon source in the frequency range from about 50 to 210~GHz. We probe millimeter wave propagation in two superconducting systems: a Josephson junction array around its plasma frequency, and a superconducting titanium film in the limit when the photon energies are larger than the superconducting energy gap. This regime is relevant for improving the performance of cryogenic quantum devices but is hard to access with conventional techniques. We show that relatively simple models can be used to describe the essential properties of the studied components.
%
%Valid PACS numbers may be entered using the \verb+\pacs{#1}+ command.
\end{abstract}


\section{Introduction}

Interactions between photons and superconducting quantum systems are of two-fold interest: Those can be directly utilized in circuit quantum electrodynamics~\cite{Wallraff2004,Blais2004} or, on the other hand, background photon emission e.g. from the warmer parts of the cryostat can be detrimental to the operation of quantum devices~\cite{Saira2012,Barends2011,Corcoles2011,deVisser2012}. 

Typical energy level spacing in a superconducting quantum device corresponds to the energy of a microwave or millimeter wave photon, which is significantly larger than thermal energy $k_BT$ at temperatures below 100~mK. The lowest microwave background levels have been obtained in experiments that combine on-chip filters with a low-noise sample stage having efficient radiation shielding and signal-line filtering~\cite{Saira2012,Kemppinen2011}. A variety of different on-chip filter elements have been used, including ground planes~\cite{Steinbach2001,Pekola2010}, thin-film resistors~\cite{Paalanen2000,Lotkhov2009,Lotkhov2012}, and superconducting quantum interference device (SQUID) arrays as tunable impedance environment~\cite{Watanabe2001,Corlevi2006} and non-uniform complex arrays as filters~\cite{Oppenlander2000,Haussler2001}.

We have used the SINIS [S$=$superconductor (Al), I$=$insulator (AlO$_x$), N$=$normal metal (AuPd))] trap as an on-chip millimeter wave detector to study spectral transmission of two stages of on-chip filters: a Josephson array filter (JAF) and a resistive capacitively shunted transmission line filter (RCF). Since millimeter waves can generate a significant amount of quasiparticles (QPs) in both studied systems, thus altering their response, it is beneficial to study them at the lowest possible signal levels that can be reached in cryogenic experiments. Our signal source is simply the electromagnetic background of a sample chamber with efficient signal-line filtering and two nested shields~ \cite{Saira2012,Barends2011,Kemppinen2011}, and we measure the rate of events caused by millimeter wave photons transmitted through the filters and absorbed by an on-chip detector.
   
\section{\label{results}Experimental methods}

A simplified schematic of the measurement circuitry is presented in Fig.~1b. At room temperature, we used floating digital-to-analog converters with a resistor or a voltage divider for the current and voltage sources, respectively. The output voltage was amplified with a battery-powered differential amplifier and measured with a digital multimeter. The cryostat body was used as the measurement ground.

\begin{figure}[h]
\center\includegraphics[width=0.75\textwidth]{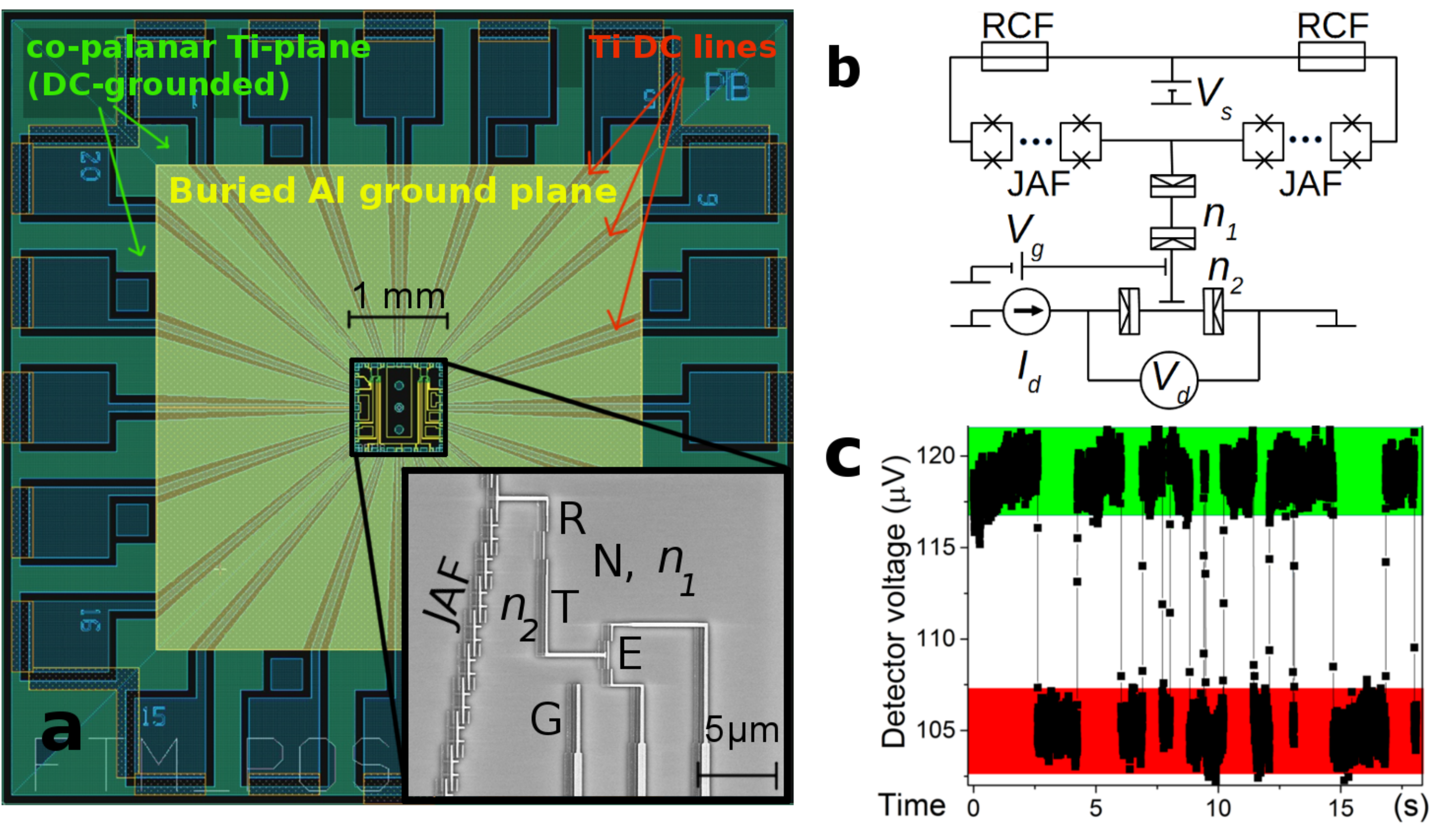}
\caption{\label{figure1} a) Overview image of the sample with the hybrid titanium transmission line filter (RCF) with buried groundplane. Inset: A scanning electron micrograph of the Josephson array filter (JAF) and the SINIS trap detector. Millimeter wave photons cause transport of individual electrons between the superconducting reservoir electrode (R) and trap node (T) via the normal metal island (N) of the SINIS single-electron trap. The change of the charge state ($n_2$) of T is detected by a SINIS single-electron transistor that acts as an electrometer (E).  b) A scheme of the device and the measurement circuit.  c) Typical two-state single-electron charge fluctuation of the SINIS trap, when the charge states $n_2=n$ (red) and $n_2=n+1$ (green) are degenerate. The trace shows the output voltage $V_\mathrm{d}$ of the SINIS electrometer that is biased with a current $I_\mathrm{d} =$ 10~pA at $T=32$ mK.}
\end{figure}

Inside the cryostat, between the room and base temperatures, flexible and resistive coaxial cables were used as signal lines. The base temperature assembly, modified from that used in Refs.~\cite{Saira2012,Kemppinen2011}, had two nested microwave shields. The outer shield of the sample stage, surrounded by the still flange shield at 800~mK, was a $\diameter 14$~cm copper cylinder mounted to a gold-plated mixing chamber flange by a thread. Bronze powder filters based on Ref.~\cite{Lukashenko2008}, screwed to the flange, were used as rf-tight signal-line feedthroughs to the outer shield. The inner shield was a $\diameter 12$~cm gold-plated copper cylinder that also had a thread for mounting a gold-plated copper cover. Resistive miniature stainless steel coaxial cables (16~cm long, $\diameter 0.33$~mm) were used as high frequency filters~\cite{Zorin1995} and signal-line feedthroughs to the inner shield. The insides of both shields were covered with microwave absorbers to prevent cavity resonances (outer shield: copper powder and Eccosorb{\textregistered}, inner shield: bronze powder). 

The detector chip was mounted on a separate copper sample inset with a printed circuit board for bonding and connectors for signal lines. The inset was attached to the sample stage with screws. These simple insets can be easily modified to suit the needs of different experiments.

The sample was fabricated on thermally oxidized silicon substrate. Using standard UV-lithography, a 50 nm thick Al groundplane was deposited and then buried under 200 nm of thermal PECVD grown SiO$_2$ (Fig.~1a). On top of the groundplane, a 20 nm thick Ti film was deposited to form the RCF and bonding pads. The superconducting and hybrid structures were fabricated with high resolution and hard Ge-mask e-beam lithography. Using three angle evaporation both the detector and the JAF were made within single lithography cycle. First, an 18 nm thick aluminium layer was deposited in UHV conditions and oxidized at low residual oxygen pressure of about 1 Pa for 10 minutes. Next another 18 nm thick aluminium layer was deposited at a different angle to form the counter electrode layer for the SQUIDs of the JAF. After the second aluminium deposition, the aluminium structure was more heavily oxidized in residual oxygen pressure of about 25 Pa for 20 minutes. As the last step, a bilayer of 3~nm of Al and 30~nm of AuPd was deposited to form the normal metal islands for the SINIS structures. The thin aluminum layer was used to enhance lattice matching to Al-AlOx electrodes.       

\section{\label{detector}Photon detector}

Even as all quantum devices are extremely sensitive to environment fluctuations, the effects of microwave photons are rarely straightforward. Radiation shielding and filtering, or lack thereof, can be observed indirectly by any quantum device, but generally this does not provide quantitative information. Here with aid of sophisticated but well established data analysis we demonstrate the possibility to utilize a SINIS single-electron trap (Fig. 1a) as a tunable barrier single-photon detector~\cite{Lotkhov2012,Lotkhov2011,Lotkhov2016} and a tool for very low signal level spectrum analysis. For the trap, photon-assisted tunneling~\cite{Pekola2010,Ingold1992,Marco2013} is the dominant electron transport mechanism, since both single- and two-electron inelastic tunneling processes can be efficiently eliminated by the Bardeen-Cooper-Schrieffer (BCS) energy gap and Coulomb blockade~\cite{Averin2008,Pekola2013}.

When a photon absorption occurs, the charge state of the normal metal island of the SINIS single-electron trap, $n_1$ (Figs.~1a~and~1b), first changes by one but relaxes back to the original state by tunneling much faster than our measurement bandwidth permits to detect. Therefore, we effectively see electron tunneling between the trap node and the superconducting reservoir electrode. 
The charge state of the trap is measured with a capacitively-coupled SINIS single-electron transistor functioning as an electrometer~\cite{Lotkhov2011}. Voltages $V_s$ and $V_g$ are applied to the control electrodes to tune the potentials of the superconducting trap node and the normal-metal island of the SINIS trap so that two charge states of the trap node, $n_2=n,n+1$, are degenerate, but a change of $n_1$ requires extra energy from a photon. The relaxation of $n_1$ can occur in either direction with equal probability, so that half of the photon absorption events change $n_2$ and can be detected.

A significant benefit of the SINIS trap as a microwave photon detector is that the threshold energy of photon assisted tunneling can be tuned between $\delta E = \Delta_\mathrm{Al}$ and $\delta E = \Delta_\mathrm{Al} + E_c$, where $\Delta_{\mathrm{Al}}\approx220~\mu$eV and $E_\mathrm{c}\approx650~\mu$eV are the BCS gap of aluminum and the charging energy of the island, 
respectively. The corresponding range for threshold frequencies $f_\mathrm{th}=\delta E /h$ is between about 50~GHz and 210~GHz ($h$ is the Planck constant). The overall quantum efficiency is low because of inevitable impedance mismatch between the high impedance tunnel junctions and the environment. 

Figure 1b shows typical time trace data of the electrometer is shown in Fig.~1c. It was symmetrically current biased with $I_d$ between 5 to 20~pA, and the resulting voltage $V_d$ was measured with a differential voltage amplifier and a digital voltmeter. Utilizing only small detector currents limits the generation of excess random telegraph noise which could be a source of significant error signals in the measurement of the microwave background~\cite{Saira2012}. The measured time constant of the electrometer, about 1.7~ms, is limited by its resistance and the capacitance of the measurement lines. The effect of missed events on the measured electron hold times $\tau$ due to the bandwidth was corrected according to Ref.~\cite{Naaman2006a}. At low frequencies, the measurement band is limited by the background charge stability and the time required to obtain sufficient statistics. In practice, we are limited to $\tau \lesssim 100$~s, but hold times longer than half an hour were observed with the highest threshold energies.

To obtain spectral information, the photon assisted tunneling (PAT) rate $\Gamma$ and thus the electron hold time $\tau$ of the trap node needs to be attained in terms of background noise spectrum. For this purpose the probability density theory was used~\cite{Ingold1992}. The tunneling rate through a NIS junction is
\begin{eqnarray}
	\label{eq:trate}
	\Gamma(\delta E) =  & 1/\tau(\delta E) =  {1 \over e^2R_T} \times  \\  
 &\int_{-\infty}^\infty dEdE^\prime n_S(E)f(T,E)  [1-f(T,E^\prime + \delta E)] P(E-E^\prime)  \nonumber,
	\label{eq:Trate}
\end{eqnarray} 
where $e$ is the elementary charge, $T$ is the electron temperature, $R_T$ is the tunneling resistance of the junction, $n_S(E)$ is the BCS density of states, $f(T,E)$ is the Fermi function, and $P(E)$ is the probability density to emit or absorb energy $E$ in the tunneling process. The effect of the charging energy is described by the electrostatic energy change $\delta E$. For simplicity, this equation is expressed for tunneling from the superconductor to the normal metal, but the opposite process can be described in a similar way.

Neglecting quantum fluctuations, $P(E)$ can be written in terms of the spectral noise density at the detector, $S_{V,\mathrm{det}}(\omega)$, as~\cite{Martinis1993}
\begin{eqnarray}
 	 	P(E)  = & \frac{1}{2 \pi \hbar} \int_{-\infty}^\infty \md t \exp \left( \frac{iEt}{\hbar}\right) \times \\
 & \exp \left( \frac{2\pi}{R_q \hbar} \int_{0}^\infty \md\omega\frac{S_{V,\mathrm{det}}(\omega)}{\omega^2} \cos (\omega t-1) \right) \nonumber,
 \end{eqnarray}
where $R_q = h/e^2$ is the resistance quantum, $\hbar =h/(2\pi)$, and $\omega = 2\pi f$ is the angular frequency of the noise. In the weak noise limit, the second exponential term can be approximated with first order expansion which is valid at energies above about $k_BT$~\cite{Martinis1993}. To cover the whole energy range, we approximate $P(E)$ with a Dirac delta function at energies below the boundary value $\epsilon_0 \sim k_BT$, and use a piecewise model 
 \begin{equation}
	\label{eq:P(E)}
	P(E) =
	\left\{
		\begin{array}{ll}
			\alpha\delta(E)  & \mbox{, } E < \epsilon_0 \\
			{\pi S_{V,\mathrm{det}}(E/\hbar) \over R_q E^2} & \mbox{, } E \geq \epsilon_0
		\end{array}
	\right.
 \end{equation}
where $\alpha$ is a normalization factor close to one. %MITEN $\epsilon_0$:N ARVO VAIKUTTAA?

\section{\label{source}Photon source}

As a source of millimeter waves in our experiment we used thermal radiation from warmer parts of the cryostat. It is a major source of microwave background for cryogenic environments at $T < 100$~mK. Its effect on quantum electronic devices is often modelled using fluctuation-dissipation theorem with an electromagnetic environment, e.g. a resistor, at an elevated temperature~\cite{Saira2012,Pekola2010,Lotkhov2011,Hergenrother1995,Saira2010,Covington2000}. We have used a slightly different approach and modelled the spectral noise density that couples to the chip, $S_{V,\mathrm{in}}(\omega)$, using Planck radiation law for a black body at elevated temperature. For our experiment this model is virtually indistinguishable from the fluctuation-dissipation model with a resistor at elevated temperature. In the sensitive range of our detector, both models yield almost identical noise spectra with only a small difference in the temperature of the radiating object. 

Starting from the Planck radiation law, we write the spectral voltage noise density, originating from the photon bath in the sample chamber, as
 \begin{equation}
        \label{eq:S_V}
 	S_{V,\mathrm{in}} = \beta_0 S_P(\omega) = \beta_0 {\hbar \omega^3 \over 4 \pi^3 c^2 } {1 \over \exp({\hbar \omega \over k_B T_P})-1}.
 \end{equation}
Here $S_P(\omega)$ is the power spectrum of the microwave radiation observed at the chip, $c$ is the speed of light, $k_B$ is the Boltzmann constant, and $T_P$ is the temperature of  the black body emitter of the model. The coupling coefficient $\beta_0$ (unit $\Omega$m$^2$) is between the black body source and the chip and can be considered frequency-independent for the sensitive frequency range of the detector. In principle $\beta_0$ could be used to compare the shielding and filtering levels of cryogenic setups even though its origin cannot be fully distinguished. However it would require identical measurement environment for the sample, for example bond wire configuration and inner sample chamber geometry. 

One should note that even though both the fluctuation-dissipation theorem and the Planck radiation law have exact physical origins, in this context they should be considered as phenomenological models. Such thermal distributions are only a crude approximation as the signal at the sample chamber has already penetrated the sample stage shields and/or signal line filters. Also in a realistic case the black body emitter temperature should have distribution based on the cryostat construction, i.e. the noise is not emitted by objects with the same temperature. 

Yet the collection of data (see Fig.~4.4 of Ref. \cite{SairaPhD2013}) from multiple experiments indicate roughly exponential decrease of the microwave background as a function of frequency~\cite{Saira2012,Pekola2010,Lotkhov2011,Covington2000}. We find the Planck model suitably simple with only two fitting parameters, of which $\beta_0$ affects only the overall magnitude of the noise, and $T_P$ also the slope of the exponential decrease. It is important also to understand the phenomenological nature of $T_P$. Both the sample chamber radiation shields, radiation coupling to the chip and the on-chip transmission lines (see Sec.~\ref{analysis}) between bonding pads and the detector affect the frequency dependence of the noise, and thus the effective $T_P$ seen by the detector.

\section{\label{results}Experimental results}

\begin{figure}[h]
\center\includegraphics[width=0.8\textwidth]{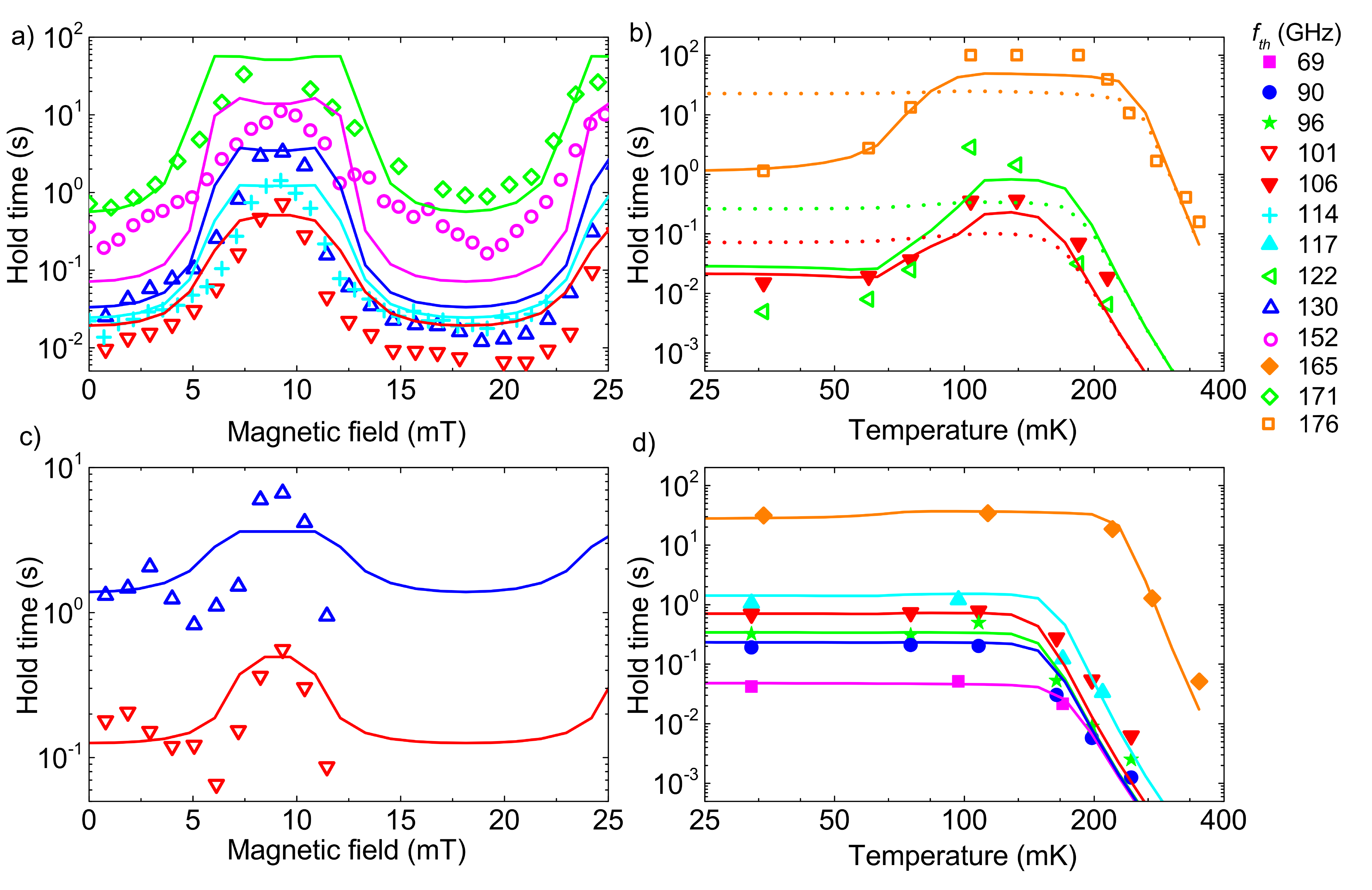}
\caption{\label{figure2} Experimental results for the electron hold times of the SINIS trap in different temperatures and magnetic fields.  Different symbols correspond to different values of $f_{\mathrm{th}}$ indicated in the legend. In each panel, the solid and dotted lines are simulation results based on $P(E)$ theory and circuit models (see the text).  a) Magnetic field dependence at base temperature $T=32$~mK.  b) Temperature dependence at $B=0$. c) Magnetic field dependence at elevated temperature $T =$~110~mK. d) Temperature dependence at $B = 8$~mT which corresponds to $\Phi = \Phi_0/2$ for the SQUIDs of the JAF filter. } 
\end{figure}

In our experiments, photons from the sample chamber were transmitted through the RCF and JAF filters and became absorbed in the detector, causing PAT events in the SINIS trap, see Figs.~1a and 1b. Electron hold times of the trap node were measured in different environment conditions (external magnetic field $B$ and bath temperature $T$) to determine attenuation properties of the filters. Spectral information was obtained by repeating the measurements with different values of threshold frequency of the trap $f_\mathrm{th}$, which is the minimum frequency that the photon must have to be able to cause a PAT event in the trap. The value of $f_\mathrm{th}$ can be adjusted by voltages $V_s$ and $V_g$ but not measured directly. Results of the experiment are shown in Fig.~\ref{figure2}, which also includes simulation results based on models described in Sec.~\ref{analysis}. The data of Figs.~2a and 2b were mainly used for studies of JAF and RCF, respectively, and the determination of the value of  $f_\mathrm{th}$ for different data sets was mainly based on the data of Fig.~2d.

The dependence of the hold time on magnetic field $B$ at the base temperature of the cryostat, about 32~mK, is presented in Fig.~\ref{figure2}a. Different symbols correspond to different values of $f_{\mathrm{th}}$. The observed $B$-dependent variation of the hold time of the trap between magnetic flux values $\Phi=0$~and~$\Phi= \Phi_0/2$ (corresponding to $B$ between 0 and 8 mT) is about two orders of magnitude. The Josephson inductance of the SQUIDs in the JAF is a periodic function of $B$, with a period of 16~mT that corresponds to a magnetic flux $\Phi=\Phi_0$ through the SQUID loops in our geometry, where $\Phi_0=h/(2e)$ is the magnetic flux quantum. This adjustable inductance of the JAF can explain the strong $B$-dependence of the hold time (see Sec. \ref{JAF}).  

Figure~\ref{figure2}b shows temperature dependence of the hold time at $B = 0$, which corresponds to the minimum attenuation of JAF. At temperatures above about 200~mK, the hold time decreases exponentially as a function of $T$ due to thermally excited tunneling events in the SINIS trap. At lower temperatures the hold time is determined by the frequency of the photon assisted tunneling events. A surprisingly large decrease in the hold time was observed with all studied $f_\mathrm{th}$ when temperature was decreased below about 100 mK. This is equal or very close to the superconducting transition temperature $T_c$ of the RCF, which is a resistive transmission line in which titanium film with $T_c \simeq 110$~mK is used as resistive material. 

The threshold frequencies were much higher than the pair-breaking frequency of titanium, $2\Delta_{\mathrm{Ti}}/h \sim 34$~GHz, where $\Delta_{\mathrm{Ti}} = 85~\mu$eV is the BCS energy gap of titanium, and thus the attenuation of RCF was expected to have only weak temperature dependence. On the other hand, we expect no temperature dependence of the JAF below $T=100$~mK, because it is extremely difficult to suppress the QP density of superconducting Al structures below the effective temperature of about 100~mK~\cite{Saira2012}. In section \ref{RCF} we show that this unexpected enhancement of high frequency propagation can be described if we consider the film as weakly connected superconducting grains.  

Figure \ref{figure2}c demonstrates that at $T = 110$~mK, the $B$-dependent variation of the hold time (factor of 4) is much smaller than at $T = 32$~mK (factor of 100). This is not surprising since RCF is a much more effective filter at $T = 110$~mK than at lower temperatures (see Fig.~\ref{figure2}b), and other effects, especially noise photons that bypass both filters, become significant. 

Finally, Fig.~\ref{figure2}d shows the temperature dependence of the hold time at $B=8$~mT. In this case, the JAF is an efficient filter and tunneling events are dominated by noise photons that bybass the filters, and hence the effect of RCF on the hold time is very small or negligible. In the PAT-dominated low temperature limit, the hold time is almost $T$-independent  for all measured $f_\mathrm{th}$. When temperature is sufficiently high to cause thermally excited tunneling in the trap, an exponential decrease of the hold time is observed. This decrease is independent of microwave background noise, and these data are used in Sec.~\ref{analysis} to determine the values of $\delta E$ and $f_\mathrm{th}$ for each set of measurements.
 
\section{\label{analysis}Analysis}

\begin{figure}[h]
\center\includegraphics[width=0.6\textwidth]{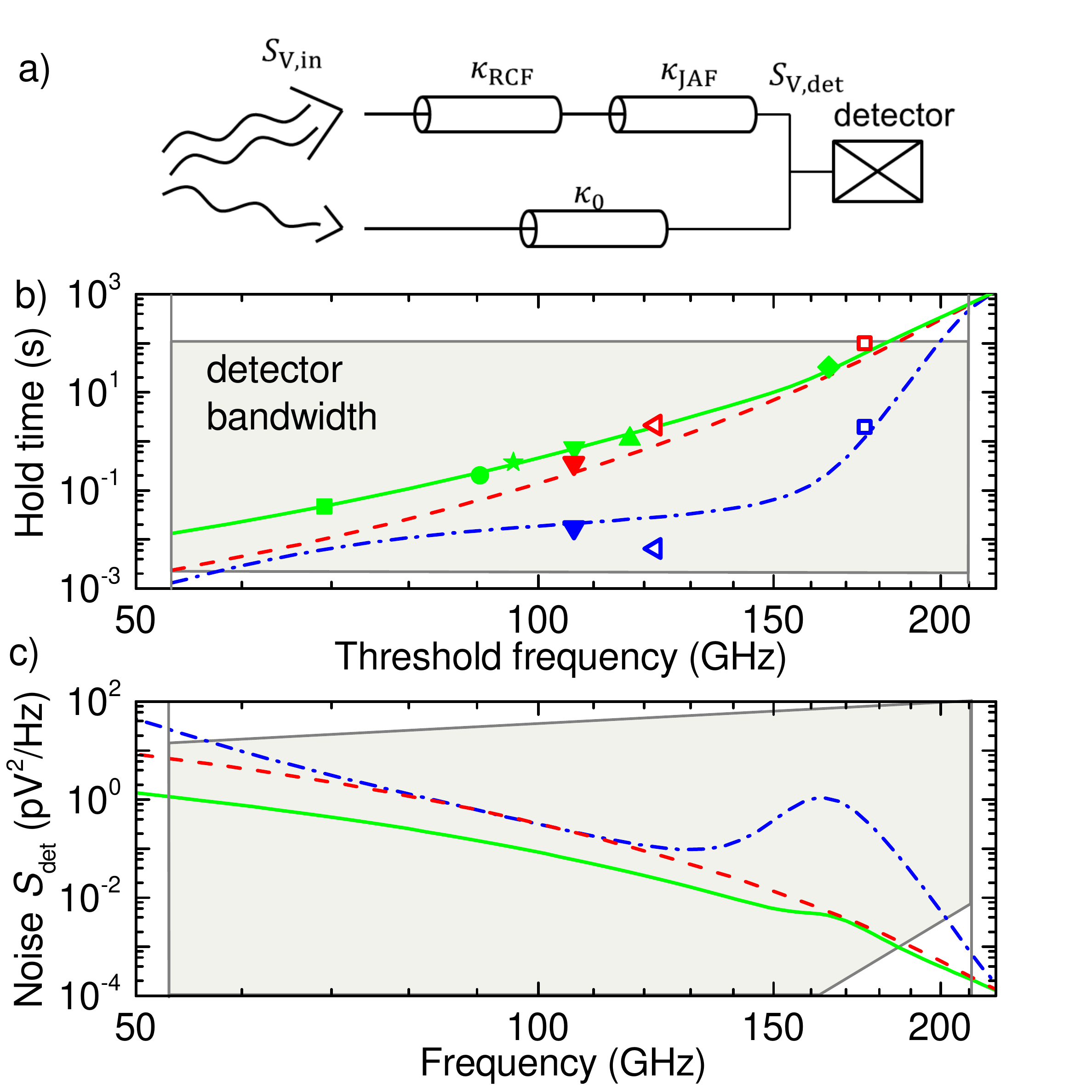}
\caption{\label{figure3} a) Model for noise propagation in the circuit. Microwave photons of the sample chamber cause voltage noise at the chip $S_{V,\mathrm{in}}$. It is transmitted to the detector via two parallel channels, through the series filter elements JAF and RCF or through electrodes near the detector without ground plane ($\kappa_0$). b) The relation between the electron hold time $\tau$ and threshold frequency $f_{\mathrm{th}}$ of the SINIS detector in different environment conditions: $B = 8$~mT and $T = 32$~mK (green solid line), $B = 0$ and $T > 100$~mK (red dashed line), and $B = 0$ and $T < 50$~mK (blue dash-dotted line). The lines are based on fits of experimental data to $P(E)$ theory, modelling $S_{V,\mathrm{det}}(\omega)$ in different experimental conditions using Planck radiation and the coupling scheme of (a). The measured hold time data are marked with corresponding symbols to Fig.~\ref{figure2}. c) Spectral density of microwave radiation reaching the detector $S_{V,\mathrm{det}}$ in different experimental conditions using the same notation as in the top panel. The sensitive range of our SINIS detector is indicated with gray shading.}  
\end{figure}

Our experimental data presented in Fig.~\ref{figure2} consist of electron hold times $\tau$ measured in different temperatures and magnetic fields. In data analysis, $\tau=1/\Gamma$ is calculated using Eq.~(\ref{eq:trate}) with the approximate form of Eq.~(\ref{eq:P(E)}) for $P(E)$. The detector is sensitive to voltage noise $S_{V,\mathrm{det}} (\omega)$ between its trap node and the superconducting reservoir electrode (see Fig.~\ref{figure1}a). $S_{V,\mathrm{det}} (\omega)$ in Eq.~(\ref{eq:P(E)}) is obtained from the spectral noise voltage picked up by the chip, $S_{V,\mathrm{in}} (\omega)$ (Eq.~(\ref{eq:S_V})), using the circuit model described in Fig.~\ref{figure3}a. The noise is filtered by the $B$-dependent JAF (transmission coefficient $\kappa_\mathrm{JAF}(B,\omega)$) and the $T$-dependent RCF (transmission coefficient $\kappa_\mathrm{RCF}(T,\omega)$) connected in series. The total transmission through these series filter elements $\kappa_F(T,B,\omega) = \kappa_\mathrm{RCF}(T,\omega) \kappa_\mathrm{JAF}(B,\omega)$ depends on temperature, magnetic field, and frequency. The models for the filter elements RCF and JAF are discussed in separate sections, Sec. \ref{RCF} and Sec. \ref{JAF}, respectively. 

When the filtering is efficient, i.e.~$\kappa_F$ is small, the noise coupled by stray capacitances from other nearby electrodes, $\kappa_0$ (see Fig.~\ref{figure3}a), dominates. We first tried using frequency independent $\kappa_0$, but noticed that then the fits to the experimental data would have required different values of $T_P$ for the noise coupled to the input of the RCF filter and for that coupled directly to the detector. A plausible explanation is that on-chip transmission lines change the frequency response in the following way. 

All signal lines coming to the sample region have essentially similar transmission line filters as the RCF that is connected to the JAF and the trap, see Fig.~\ref{figure1}a. However, there is a 1~mm~$\times$~1~mm region around the detector where each line continues about the distance $d\approx 0.5$~mm without the ground plane. We assume that the noise bypassing the filters is dominated by microwave photons entering this region, and we model its coupling to the detector with a lumped element RC transmission line. 

We calculate frequency-dependent transmission coefficients $\kappa(\omega)$ for lossy lumped element transmission lines using Telegraphers equations, see for example~\cite{Pozar1998}. Only dissipative losses of the conductor are taken into account as those are considered to be the dominating signal attenuation mechanism for our filters, and we write
\begin{equation}
	\label{eq:kappa}
	\kappa(\omega)=V_{\mathrm{out}}(\omega)/V_{\mathrm{in}}(\omega)=e^{-\mathrm{Re}[\gamma(\omega)] l},
\end{equation}
where $V_{\mathrm{in}}$ and $V_{\mathrm{out}}$ are voltages in the input and output ports of the filter, respectively, $l$ is the length of the transmission line, and ${\gamma=\sqrt{i\omega C'_\mathrm{shunt} Z'_\mathrm{series}}}$ is the complex propagation coefficient, where $Z'_\mathrm{series}$ is the series impedance and $C'_\mathrm{shunt}$ the shunt capacitance per unit length for the tranmission line.  

Using Eq.~(\ref{eq:kappa}) we obtain the transmission coefficient for the signal lines without the ground plane $\kappa_0 = A  \exp(-\mathrm{Re} [\sqrt{i\omega C'R'}]d)$. Coupling of the microwave background noise into the signal lines is weaker in the middle of the chip than on the bonding pads outside the ground plane, and that difference is quantified by the fitting parameter $A<1$. In the frequency range of our experiments, the product $\kappa_0S_{V,\mathrm{in}}$ has a shape very close to the Planck spectrum of Eq.~(\ref{eq:S_V}) but with a different black body temperature $T_P$. The best fits to the data are obtained using realistic parameters $R' = 4.5~\Omega$/$\mu$m and $C' = 300$~aF/$\mu$m for resistance and capacitance per length, respectively, and $A=0.1$. 

The relation between voltage noise density entering the chip, $S_{V,\mathrm{in}}$ in Eq.~(\ref{eq:S_V}), and reaching the detector, $S_{V,\mathrm{det}}$, is described by the transmission coefficient $\kappa(T,B,\omega)$:
 \begin{equation}
        \label{eq:kokonaismalli}
S_{V,\mathrm{det}} = \kappa_\mathrm{tot} S_{V,\mathrm{in}} =[\kappa_\mathrm{JAF}(B,\omega) \kappa_\mathrm{RCF}(T,\omega) + \kappa_0(\omega)] S_{V,\mathrm{in}}.
\end{equation}
Even though the simulations of Fig.~\ref{figure2} are based on this relatively complicated overall transmission coefficient (\ref{eq:kokonaismalli}), we note that $\kappa_\mathrm{JAF}$, $\kappa_\mathrm{RCF}$, and $\kappa_0$ each depend mainly on a single set of measurements: Figs.~\ref{figure2}a, \ref{figure2}b, and \ref{figure2}d, respectively. This helped us to ensure that unambiguous values were obtained for the fitting parameters.

The starting point for the data analysis are the results of Fig.~\ref{figure2}d, which were measured when the JAF had large attenuation and the PAT effects of the detector were dominated by the noise that had bypassed the filters via the transmission channel $\kappa_0$. The measurements were done with 6 values of $f_{\mathrm{th}}$, whose value for each data set was determined from fits to the data. In the first iteration, the transmission through the filters was not taken into account at all, but the final results presented as solid lines in Fig.~\ref{figure2}d are calculated with the full model of Fig.~\ref{figure3}a, including the models for the RCF and JAF. 

On second iteration we simultaneously fitted all data sets with the Eq. (\ref{eq:kokonaismalli}). The common fitting parameters $T_P=1.5$~K and $\beta_0=4\times10^{-2}$~$\Omega$m$^2$ were obtained from these fits. The datasets  extending from the low- to the high-temperature limit, where PAT and thermally excited tunnelling dominate, respectively, were used to define $f_\mathrm{th}$ for each dataset of Fig.~2.  

The green line and green symbols of Fig.~\ref{figure3}b show the relation between the low-temperature electron hold time and threshold frequency $f_{\mathrm{th}}$ of the trap for the experimental case of Fig.~\ref{figure2}d. Fig.~\ref{figure3}c shows the corresponding spectral noise density near the detector as a function of frequency. The other lines in Figs.~\ref{figure3}b and \ref{figure3}c correspond to experimental conditions where the on-chip filters are effective, and they have been calculated using Eq.~(\ref{eq:trate}) with Eqs.~(\ref{eq:P(E)}) and (\ref{eq:S_V}), and applying the filter models of Sections \ref{RCF} and \ref{JAF} in Eq. ~(\ref{eq:kokonaismalli}). The sensitive range of our SINIS detector is shaded with gray.

\subsection{\label{RCF}Resisitive transmission line}

As shown in the large-scale image of Fig.~1a, the resistive transmission line filters (RCF) consists of a titanium electrode which is capacitively coupled to a buried aluminum ground plane. The length of the filter, about 3.5 mm, is comparable to the wavelength of microwaves in the frequency range of our experiments. The Ti layer is highly resistive ($\sim 100$~$\Omega/\square$) in the normal state, i.e.~when either temperature or magnetic field is above the critical value of about 110~mK or 40~mT, respectively. The microwave attenuation of RCF was not expected to change much at the superconducting transition of Ti, but the experiments of Fig.~\ref{figure3}b show that the hold time of electrons in the SINIS detector increases by two orders of magnitude when temperature decreases below $T_c$ of titanium film, about 110 mK.

\begin{figure}[h]
\center\includegraphics[width=0.6\textwidth]{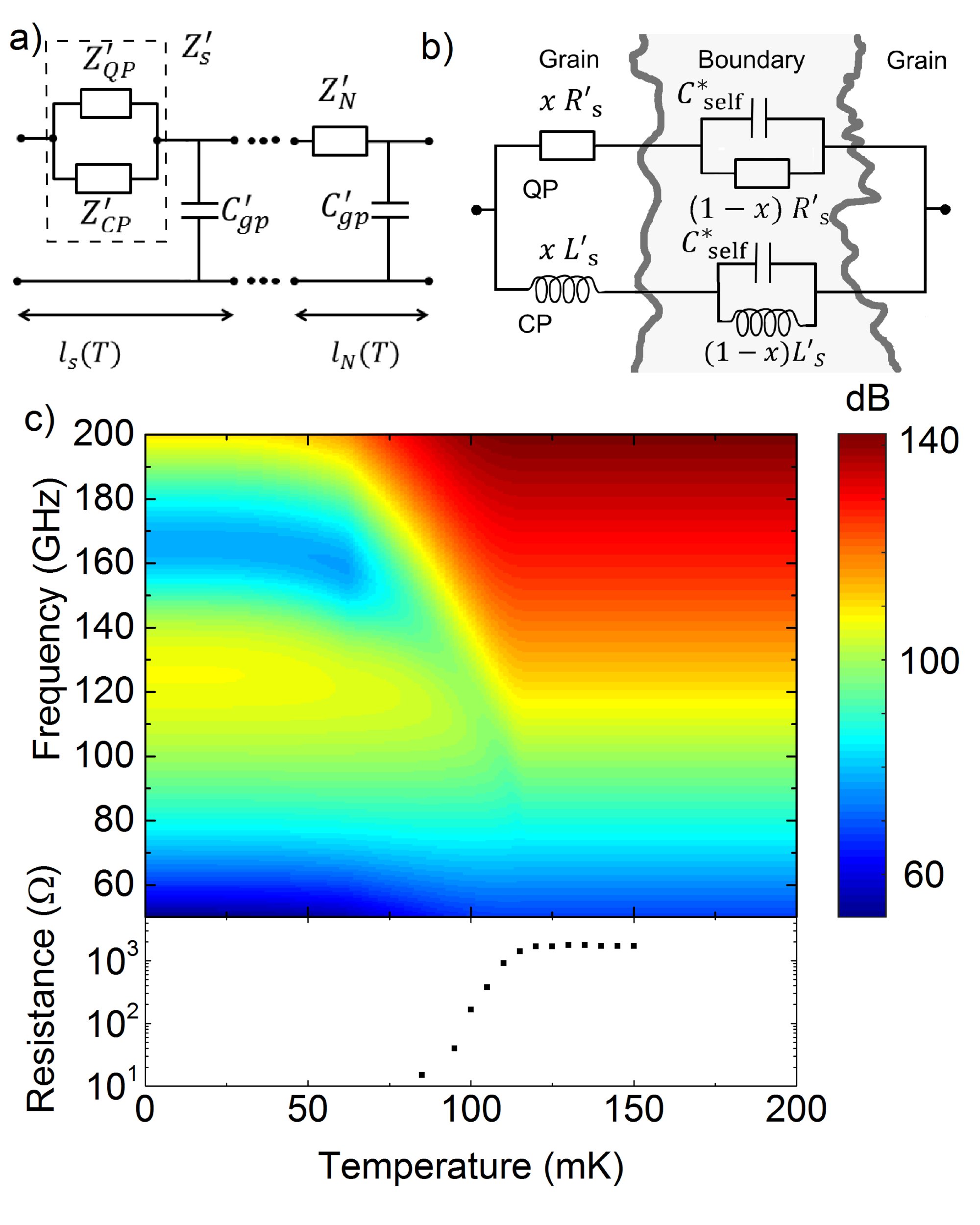}
\caption{\label{figure4} a) The circuit model used for the RCF consisting of the superconducting (left) and normal-metal parts (right). b) Circuit model for superconducting state impedance per unit length $Z'_S$ for the film with consisting of weakly connected grains. c) Top panel shows attenuation $1/\kappa_\mathrm{RCF}$ in decibel scale as a function of temperature and frequency for the RCF modelled as weakly connected grains as explained in the text. The bottom panel shows the observed dc-resistance of the RCF as a function temperature.}  
\end{figure}

The superconducting transition of a thin nonhomogeneous film has a finite width as a function of temperature due to local variations of $T_c$ and phase fluctuations of the superconducting order parameter. In the model this was taken into account by dividing the transmission line to superconducting and normal-metal parts connected in series, with temperature dependent lengths $l_{\mathrm{S}}(T)$ and $l_{\mathrm{N}}(T)$, respectively, as shown in Fig.~\ref{figure4}a. Temperature dependencies of $l_{\mathrm{S}}(T)$ and $l_{\mathrm{N}}(T)$ relative to the total length $l=3.5$~mm of the transmission line were obtained from the temperature dependence of the dc resistance of the RCF near $T_c$ of Ti film shown in the bottom panel of Fig.~\ref{figure4}c. 

To model the high frequency conductivity in the superconducting film well established theory formulated by Mattis and Bardeen~\cite{Mattis1958} was utilized. Frequency dependent complex conductivity of a superconductor can be written as $\sigma_\mathrm{MB}(\omega, T)~=~\sigma_\mathrm{MB1}(\omega, T)~-~i\sigma_\mathrm{MB2}(\omega, T)$, where
	\begin{eqnarray}
 	{\sigma_\mathrm{MB1} \over \sigma_N} = &{2 \over \hbar \omega} \int^{\infty}_{\Delta_{\mathrm{Ti}}} {[f(E) - f(E + \hbar\omega)](E^2+\Delta^2_{\mathrm{Ti}} + \hbar \omega E) \over  \sqrt{E^2 - \Delta^2_{\mathrm{Ti}}} \sqrt{(E+ \hbar \omega)^2 - \Delta^2_{\mathrm{Ti}}}} dE \\
 	&+ {1 \over \hbar \omega} \int^{-\Delta_{\mathrm{Ti}}}_{\Delta_{\mathrm{Ti}}-\hbar \omega} {[1- 2f(E + \hbar \omega) ](E^2+\Delta^2_{\mathrm{Ti}} + \hbar \omega E) \over  \sqrt{E^2 - \Delta^2_{\mathrm{Ti}}} \sqrt{(E+ 		\hbar \omega)^2 - \Delta^2_{\mathrm{Ti}}}} dE \nonumber
 	\end{eqnarray} 
is a dissipative part with thermal and photon excited QPs and 
\begin{equation}
{\sigma_\mathrm{MB2} \over \sigma_N} = {1 \over \hbar \omega} \int^{\Delta_{\mathrm{Ti}}}_{-\Delta_{\mathrm{Ti}}} {[1- 2f(E + \hbar \omega) ](E^2+\Delta^2_{\mathrm{Ti}} + \hbar \omega E) \over  \sqrt{E^2 - \Delta^2_{\mathrm{Ti}}} \sqrt{(E+ \hbar \omega)^2 - \Delta^2_{\mathrm{Ti}}}} dE
\end{equation}
is the dissipationless superfluid response. In the equations above, $\sigma_N$ is the normal-state conductivity of the titanium film.

First we demonstrate that the temperature dependence of the RCF cannot be explained solely by the superconductivity of the film. We used the circuit of Fig.~\ref{figure4}a and a finite length lumped element RLC transmission line model~\cite{Pozar1998} to calculate the transmission coefficient (Eq.~(\ref{eq:kappa})) for the RCF 
\begin{eqnarray}
\label{eq:RCF}
\kappa_\mathrm{RCF}(T,\omega) = &\exp \Big{(}-\mathrm{Re} \big{[}\sqrt{(i\omega C'_\mathrm{gp} Z'_N(T,\omega)}\big{]} l_{\mathrm{N}}(T) \\ 
&-\mathrm{Re}\big{[}\sqrt{(i\omega C'_\mathrm{gp} Z'_S(T,\omega)}\big{]}l_{\mathrm{S}}(T) \Big{)} \nonumber,
\end{eqnarray}
where $Z'_N = R'_N$ is the normal state and $Z'_{S} = {Z'_{QP} Z'_{CP} / (Z'_{QP} + Z'_{CP})}$  
is the superconducting state impedance per unit length and $Z'_{QP} = R'_\mathrm{QP} = R'_\mathrm{N}/(\sigma_\mathrm{MB1}/\sigma_N)$ is the impedance per unit length for the QP channel and $Z'_{CP} = i\omega L'_\mathrm{CP} = i \omega (R'_{N}/\omega)/(\sigma_\mathrm{MB2}/\sigma_N)$ for the Cooper pair (CP) channel. The normal state resistance of the Ti film per unit length $R'_{N}=505~\Omega$/mm is  obtained from an independent measurement. The capacitance to ground plane was used as fitting parameter and the best fits were obtained with $C'_\mathrm{gp}=0.2$~pF/mm, which is in reasonable agreement with the value of 1~pF/mm estimated from geometry. Equation (\ref{eq:RCF}) was applied in Eqs. (\ref{eq:kokonaismalli}), (\ref{eq:P(E)}) and (\ref{eq:trate}) to simulate the electron hold time of the detector as a function of temperature, assuming that JAF is temperature-independent below ${\sim 110}$~mK. The dotted lines in Fig. 2b are results of this model. As can be seen, this model does not explain the observed decrease of the hold time at low temperatures.

Instead, the observations can be at least semi quantitatively explained by the disorder of the Ti film evaporated in the presence of residual oxygen. We assume that the Ti film consists of weakly connected grains that give rise to a significant self-capacitance. We modelled the capacitive coupling between the randomly distributed grains using a normalised Gaussian distribution $\mathcal{N}({C}^*_\mathrm{self}|\mu ,\sigma)$, where the expectation value $\mu = \bar{C}^*_\mathrm{self} = 0.3$~aFmm and standard deviation $\sigma = 0.5 \mu$, were used as fitting parameters for the capacitive impedance per unit length, $Z'_C = (i \omega C^*_\mathrm{self})^{-1}$.

Figure \ref{figure4}b represents simplified circuit of the coupled superconducting grains that was used to simulate the RCF. We used the circuit in Fig.~\ref{figure4}a revised so that the self-capacitances for both normal state and superconducting impedances were taken into account:
\begin{eqnarray}
Z'_{N} =   xR'_{N} +(1-x)\int_0^\infty\mathcal{N}(C^*_\mathrm{self}|\mu ,\sigma)\left({1 \over R'_N} +  i \omega C^*_\mathrm{self}\right)^{-1} dC^*_\mathrm{self} \\
Z'_{S} =   \int_0^\infty \mathcal{N}(C^*_\mathrm{self}|\mu ,\sigma) {Z'_{QP} Z'_{CP} \over  Z'_{QP} + Z'_{CP}}  dC^*_\mathrm{self}. \\
\end{eqnarray}
Here the QP and CP channel impedances are 
\begin{eqnarray}
Z'_{QP} = x R'_S + (1-x)\left({1 \over R'_S} +  i \omega C^*_\mathrm{self}\right)^{-1} \\
Z'_{CP} = x (i\omega L'_S) + (1-x)\left({1 \over i \omega L'_S}+  i \omega C^*_\mathrm{self}\right)^{-1}.
\end{eqnarray}
The impedance ratio $x=0.1$ between the comparably large well conducting grains and thin semi-insulating grain boundaries (Fig.~\ref{figure4}b) was used as a fitting parameter. We again calculated the transmission using Eqs.~(\ref{eq:RCF}), (\ref{eq:kokonaismalli}), (\ref{eq:P(E)}) and (\ref{eq:trate}) and a reasonably good agreement with data was obtained (Fig.~2b solid lines).

The temperature and frequency dependent attenuation $1/\kappa_\mathrm{RCF}(T,\omega)$ is shown in the top panel of Fig.~4c. The effect of the self-capacitance is significant only at low temperatures, where the film is in the superconducting state. LC-resonances of the superfluid kinetic inductance and grain boundary capacitances enhance the conductivity in the frequency range from about 120 to 220 GHz. As seen in Fig.~3b this reduces the hold time for all values of $f_\mathrm{th}$ that can be measured with the bandwidth of our detector. For the QP and the normal state channels the conductivity increase is small because the impedance of the capacitive shunt is higher than the QP or the normal state resistance.     

We want to stress that a proper microscopic model would be required for better understanding of the phenomenon, which is, however, beyond the scope of this experimental paper. This simple phenomenological model yields with realistic parameters a plausible explanation for our observations of unexpected high frequency propagation in a disordered superconducting film. 

\subsection{\label{JAF}Josephson array filter}

The Josephson array filter (JAF) consists of two branches, with 5 and 35 aluminum SQUIDs in series, respectively (Fig.~1a). The effects studied in this work are dominated by the 5-SQUID branch. Its attenuation can be tuned by a magnetic field, since each SQUID of the JAF is effectively a tunable Josephson inductance element with 
\begin{equation}
	\label{eq:L_J}
	L_\mathrm{J}(\Phi) = \Phi_0 / (2\pi I_\mathrm{c}(\Phi)).
 \end{equation}
The critical current $I_{c}(\Phi) \simeq I_{c}(0) \cos(\pi\Phi / \Phi_0$) has maxima and minima when the magnetic flux $\Phi$ through the SQUID loop is integer or half integer multiple of the flux quantum $\Phi_0$, respectively. The magnetic field that corresponds to $\Phi_0$ in the SQUID loops of JAF is about 16~mT. 

\begin{figure}[h]
\center\includegraphics[width=0.6\textwidth]{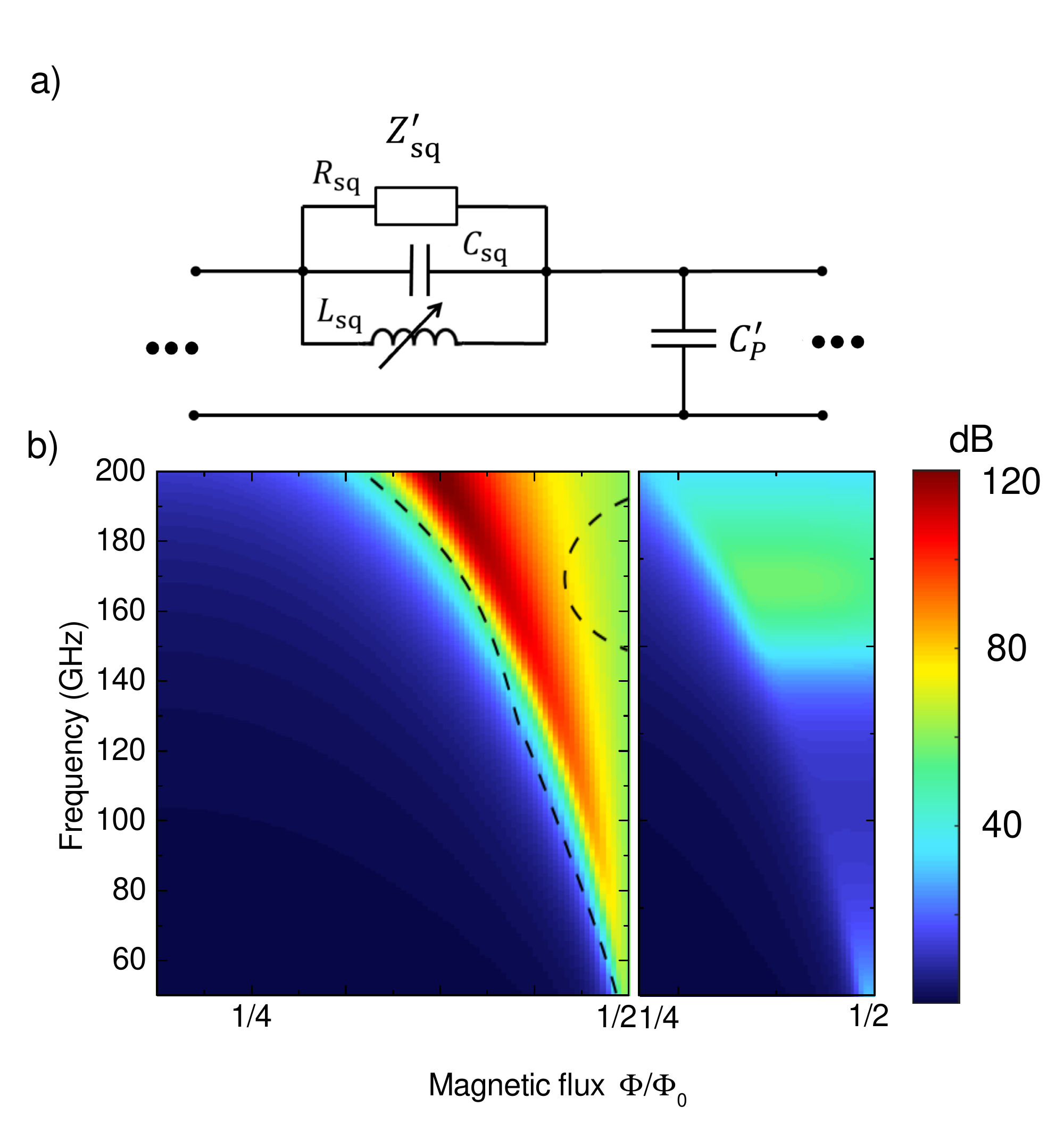}
\caption{\label{figure5} a) The transmission line element used in analysing the Josephson array filter.  b) Left panel: Attenuation $1/\kappa_\mathrm{JAF}$ calculated for the JAF model at low temperature $T \ll 100$~mK as a function of magnetic flux $\Phi$ and frequency $f$. The plot covers the range of $\Phi$ from 0 to $\Phi_0/2$, but the attenuation is a periodic function of $\Phi$ with period $\Phi=\Phi_0$, and the range from $\Phi_0/2$ to $\Phi_0$ is a mirror image of the range covered in the plot. When frequency and magnetic flux are in the region between the black dashed lines, experiments are dominated by the parasitic coupling $\kappa_0$ (see Fig. 3a) and the effects of JAF cannot be observed. Right panel: Additional attenuation of JAF when the effects of RCF and parasitic coupling are taken into account.}
\end{figure}

The frequency range of the SINIS trap detector is higher or comparable to the magnetic flux dependent plasma frequency of the SQUIDs {$f_{p}~\sim~20 - 200$~GHz}. This region is hard to model theoretically. We use a simplified model suitable for experimental work. The chain of the 5 SQUIDs of the JAF, with total length about $l_\mathrm{chain} = 15~\mu$m, is treated as a lumped element RLC-transmission line circuit~\cite{Pozar1998} shown in Fig.~3c. The model is similar to that in Ref. \cite{Haviland2000} but takes into account transmission losses to make the model applicable above the plasma frequency. Each SQUID element consists of three parallel elements, kinetic inductance $L_\mathrm{sq}(\Phi) =\Phi_0/(2\pi I_{c}(\Phi))$, total junction capacitance $C_\mathrm{sq}=0.5$~fF and resistive damping term $R_\mathrm{sq} = 10$~k$\Omega$. Thus the SQUID impedance per unit length is 
\begin{equation}
Z'_\mathrm{sq} = \Big{(}i\omega C_\mathrm{sq}+ {1 \over i\omega L_\mathrm{sq}} + {1 \over R_\mathrm{sq}}\Big{)}^{-1}/l_\mathrm{sq}, 
\end{equation}
where $l_\mathrm{sq} = 3~\mu$m is the length of one SQUID. The capacitance per unit length between the transmission line and environment is estimated to be about $C'_p = 300$~aF$/ \mu$m. The critical current of the Josephson junctions at $\Phi=0$, $I_{c}(0)=$~0.49~$\mu$A, was used as a fitting parameter. The best fits were obtained with a critical current 2.7 times the value estimated from the measured junction resistance $R_{J} = 2.3~$k$\Omega$ and superconducting energy gap {$\Delta_{\mathrm{Al}}=220~\mu$}eV using the Ambegaokar-Baratoff relation~\cite{Ambegaokar1963}. The junction capacitance was estimated from the total junction area of each SQUID $A_j = 2 \times 0.01$~$\mu$m$^2$, oxide thickness $d = 1.5$~nm, and relative permittivity $\epsilon = 4.5$.  

The damping resistance $R_\mathrm{sq}$ is considered to originate from dielectric losses and quasiparticle excitations~\cite{Martinis2005,Hassel2013}. The dielectric losses may have pronounced contribution because of the very low signal level~\cite{O'Connel2008} and high photon frequency~\cite{Skacel2015} compared to typical experiments. Using only the cryogenic microwave background as signal source allows probing the system with minimal disturbance, which maximises the density of the unexcited two level fluctuators present in dielectrics. On the other hand, the detection band photons from 50~GHz to 210~GHz certainly have enough energy to cause excitations of the fluctuators.

Transmission coefficient of the JAF, based on Eq.~(\ref{eq:kappa}) is
\begin{equation}
	\kappa_\mathrm{JAF}(B) = \exp \left[ -\mathrm{Re} [\sqrt{i\omega C'_p Z'_\mathrm{sq}}]l_\mathrm{chain} \right] , 
\end{equation}	
which depends on magnetic field via $L_\mathrm{sq}(\Phi)$. It was used in Eqs. (\ref{eq:kokonaismalli}), (\ref{eq:P(E)}) and (\ref{eq:trate}) to simulate magnetic field dependence of the hold time. Reasonably good agreement with the data in Fig.~\ref{figure2}a with all $f_\mathrm{th}$ was obtained. The calculated attenuation $1/\kappa_\mathrm{JAF}$ is presented in the left panel of Fig.~\ref{figure5}b. The maximum attenuation is not reached at $\Phi = \Phi_0/2$ but at a value of $\Phi$ corresponding the LC-resonance (plasma frequency) of the SQUID, which for higher frequencies occurs at values of $\Phi$ further from $\Phi_0/2$. This leads to the widening of the magnetic field range where the attenuation is high as $f_\mathrm{th}$ increases. The effect can be observed in the experimental results of Fig.~\ref{figure2}a. 

However, the resonances themselves cannot be observed directly in the experiment, since the parasitic coupling, $\kappa_0$, dominates when JAF has a large attenuation. The region where $\kappa_0$ dominates is marked in the left panel of Fig.~\ref{figure5}b with dashed lines. The right panel of Fig.~\ref{figure5}b demonstrates the effective filtering performance of JAF when the leakage caused by parasitic coupling $\kappa_0$ is taken into account. The ratio $\kappa_\mathrm{tot}/\kappa_\mathrm{wo-JAF}=(\kappa_\mathrm{JAF}\kappa_\mathrm{RCF}+\kappa_0) / (1 \times \kappa_\mathrm{RCF}+\kappa_0)$ is plotted as a function of magnetic flux and frequency in the low-temperature limit, where the attenuation of RCF is minimized. Here $\kappa_\mathrm{wo-JAF}$ is the total transmission coefficient when JAF is replaced by perfect transmission $\kappa = 1$. 

In spite of the simplicity of the model with respect to the physical complexity of the system, the uncertainty of the experimentally defined parameters, and the unavoidable structural non-idealities, the fits in Fig.~\ref{figure2}a are in good agreement with experimental results. The model explains both the magnitude of the $B$-dependent modulation of the hold times with different $f_{\mathrm{th}}$ and the widening of the magnetic field range of high attenuation at higher  $f_{\mathrm{th}}$. However, the model is phenomenological and the strict origin of parameters as well as the suitability of the transmission line model for the system should be considered critically. Better validation of the model requires experiments with more simplified systems, without the RCF, and where the JAF attenuation would not be limited by $\kappa_0$. 

Finally, the full coupling model of Fig.~\ref{figure3} also explains at least semi-quantitatively the result of Fig.~\ref{figure2}c, i.e., the much weaker magnetic field dependence of the hold time at $T$ = 110~mK, than at $T$ = 32~mK (Fig.~\ref{figure2}c). The reason is that at $T$ = 110~mK the attenuation of RCF is large, which means that a considerable fraction of detected photons arrive via the $B$-independent parasitic coupling $\kappa_0$ in all magnetic fields, whereas at $T = 32$~mK the attenuation of RCF is much smaller and photons arrive to the detector mainly via RCF and JAF, and $\kappa_0$ has an important contribution only in magnetic fields where the attenuation of JAF is high. 

\section{Conclusions}

We demonstrated that a single-electron trap excited by individual photons can be used to obtain spectral information of cryoelectronic devices at millimeter wave frequencies with extremely low signal levels. Our source of millimeter wave photons was the electromagnetic background in a well shielded and filtered sample chamber at temperatures between 30 and 400 mK. Our results are consistent with an attenuated Planck spectrum of a black-body emitter at 1.5 K before the filter elements. 

Two series-connected elements were used as cryogenic millimeter wave filters, a 15-$\mu$m-long Josephson array filter (JAF) consisting of an array of aluminium SQUIDs, and a 3.5-mm-long resistive transmission line filter (RCF) made of titanium. Both filters were effective attenuators for millimeter waves. When the attenuation of JAF was tuned by a magnetic field, the hold time of electrons in the single-electron trap changed by two orders of magnitude. We demonstrated that even a simple circuit model, in spite of the theoretical complexity of the SQUID at frequencies comparable to the plasma frequency, is suitable for a qualitative describtion of the millimeter wave transmission. In RCF, the attenuation was observed to decrease considerably when temperature was decreased below the superconducting transition temperature of titanium. This behaviour can be explained by resonance effects in a disordered metal film consisting of weakly connected superconducting grains.

We thank A. B. Zorin, J. Hassel, O.-P. Saira and V. F. Maisi for discussions. The research was financially supported by the Wihuri foundation (E.M.) and by the Academy of Finland (A.K., grant 259030). This work belongs to the Joint Research Project MICROPHOTON of the European Metrology Research Programme (EMRP). The EMRP is jointly funded by the EMRP participating countries within EURAMET and the European Union.
\\

\end{document}